%% file: enubet_coffani_v2.tex
\documentclass[12pt]{article}
\usepackage{graphicx}

\newcommand\pubnumber{NuPhys2017-Coffani}
\newcommand\pubdate{\today}

\def\bicocca{Universit\`a degli Studi di Milano Bicocca and INFN, \\ Piazza della Scienza 3, I-20126, Milano, Italy}
\def\support{\footnote{This project has received funding from the European Union's Horizon
2020 Research and Innovation programme under Grant Agreement
no. 654168 and no. 681647.}}

\def\Title#1{\begin{center} {\Large #1 } \end{center}}
\def\Author#1{\begin{center}{ \sc #1} \end{center}}
\def\Address#1{\begin{center}{ \it #1} \end{center}}

\newcommand\pubblock{\rightline{\begin{tabular}{l} \pubnumber\\
         \pubdate  \end{tabular}}}
\newenvironment{Abstract}{\begin{quotation}  }{\end{quotation}}
\newenvironment{Presented}{\begin{quotation} \begin{center} 
             PRESENTED AT\end{center}\bigskip 
      \begin{center}\begin{large}}{\end{large}\end{center} \end{quotation}}

\input econfmacros.tex

\begin{document}
\begin{titlepage}
\pubblock

\vfill
\Title{A narrow band neutrino beam with high precision \\ flux measurements}
\vfill
\Author{ A. Coffani\support}
\Address{\bicocca}
\Author{ on behalf of the ENUBET Collaboration: \\ 
G.~Ballerini, A.~Berra, 
R.~Boanta, M.~Bonesini, C.~Brizzolari, G.~Brunetti, M. Calviani, S. Carturan, M.G.~Catanesi, S.~Cecchini,  F.~Cindolo, 
G.~Collazuol, E.~Conti, F.~Dal Corso, G.~De Rosa, A.~Gola, R. A. Intonti, C.~Jollet, Y.~Kudenko,  M.~Laveder, A.~Longhin, P.F. Loverre, L.~Ludovici,  
L.~Magaletti, 
G.~Mandrioli,
A.~Margotti,  
V.~Mascagna,
N.~Mauri,
A.~Meregaglia,
M.~Mezzetto,
M. Nessi,
A.~Paoloni,
M.~Pari,
L.~Pasqualini,
G.~Paternoster,
L.~Patrizii,
C.~Piemonte,
M.~Pozzato,
F.~Pupilli,
M.~Prest,
E.~Radicioni, C.~Riccio,
A.C.~Ruggeri,
G.~Sirri,
M.~Soldani, 
M.~Tenti, M. Torti, 
F.~Terranova,
E.~Vallazza,
M. Vesco,
L.~Votano
}

\vfill
\begin{Abstract}
The ENUBET facility is a proposed narrow band neutrino beam where lepton production is monitored
at single particle level in the instrumented decay tunnel. This facility addresses simultaneously the two most important challenges for the next generation of cross section experiments: a superior control of the flux and flavor composition at source and a high level of tunability and precision in the selection of the energy of the outcoming neutrinos. We report here the latest results in the development and test of the instrumentation for the decay tunnel. Special emphasis is given to irradiation tests of the photo-sensors performed at INFN-LNL and CERN in 2017 and to the first application of polysiloxane-based scintillators in high energy physics.    
\end{Abstract}
\vfill

\begin{Presented}

NuPhys2017, Prospects in Neutrino Physics

Barbican Centre, London, UK,  December 20--22, 2017

\end{Presented}

\vfill
\end{titlepage}
\def\thefootnote{\fnsymbol{footnote}}
\setcounter{footnote}{0}

\section{Introduction}
The next generation of neutrino cross section experiments must
address several accelerator and detector challenges to overcome the
limitations of current measurements. In order to reach a 1\% precision
on absolute cross sections, the dominant uncertainties on flux can be
removed measuring in a direct manner the neutrino production in the
decay tunnel. In addition, a tunable, narrow band beam is needed to
constrain the neutrino energy at source  without relying on the
reconstruction of the neutrino interaction products. The ENUBET
Collaboration~\cite{enubet} is designing a facility that addresses these
challenges and enables a new generation of short baseline experiments
that will run in the DUNE and HyperKamiokande era. These experiments will impact in a
substantial manner in the systematic reduction programme of long baseline
facilities. The electron neutrino production in ENUBET is monitored at
the 1\% level through the observation of positrons in the decay tunnel
originating from the three body semileptonic decay of the charged
kaons: $K^+ \rightarrow \pi^0 e^+ \nu_e$ (and its CP-conjugate during
the antineutrino run) \cite{Longhin:2014yta}. The other leptonic and
semileptonic modes are employed to measure the $\nu_\mu$ flux at the
kaon peak and constrain the $\nu_\mu$ flux from pion decays. Results on
the beamline and the instrumentation achieved during the first year
of the project are summarized
in~\cite{Berra:2016thx,test_12module,Meregaglia:2016vxf,Ballerini:2018hus,pupilli_nufact2017}
and the updated design of the facility based on the static focusing
system will be presented in summer 2018. In this Poster we focus on the
results achieved in 2017 on the instrumentation for the decay
tunnel. Special emphasis is given to irradiation tests of the
photo-sensors performed at INFN-LNL and CERN in 2017 and the studies on
polysiloxane-based scintillators.

\section{Irradiation tests}
Positron identification and monitoring is achieved in ENUBET by means of calorimetric techniques. The instrumentation of the decay tunnel is based on a shashlik iron-plastic-scintillator calorimeter with a 4.3 X$_0$ longitudinal segmentation and a photon veto. The calorimeter is installed on the outer walls of the tunnel and hence only large angle particles - mostly from kaon decays - reach the detector before the beam dump.
The scintillator light is read out by Silicon Photomultipliers (SiPMs) directly connected to the optical fibers.  Thanks to the compactness of these photosensors, a fine-grained longitudinal segmentation can be achieved without introducing dead areas in the calorimeter.  The technology platform for the ENUBET light sensors is the RGB-HD technology developed by Fondazione Bruno Kessler (FBK).
The calorimeter is designed to be operated in the harsh environment of the decay tunnel and the SiPM located in the bulk of the calorimeter will be exposed to sizable neutron fluxes originating from hadronic showers. The integrated dose in the innermost calorimeter layers (1 m radius) is 1.9$\times$10$^{11}$ n/cm$^2$ (1 MeV-eq).  
It corresponds to 1.95 $\times$ 10$^{17}$ K$^+$ decays and $\sim$ 10$^4$ $\nu_e^{CC}$ events at a 500 ton neutrino detector located 50~m after the beam dump. 
For this reason a dedicated irradiation test has been performed at the
CN Van de Graaff irradiation facility at INFN-LNL (Legnaro) in June 2017. SiPMs of different pixel sizes were irradiated up to 10$^{12}$ n/cm$^2$ (1 MeV-eq) and their response in the ENUBET calorimeter modules were tested at the CERN East Area facility. The exposure reproduces the effects of neutron irradiation on the SiPM detectors arising from  hadronic interactions in the passive material of calorimeter. We measured both the dark current increase and the response to mip before and after the irradiation.   

\begin{figure}[htb]
\centering
\includegraphics[height=3.1in]{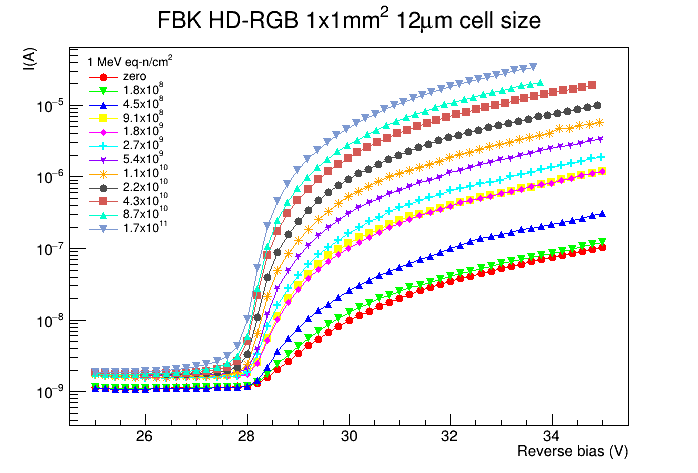}
\caption{I-V curves for 12 $\mu$m pixel SiPM at
different irradiation fluences.}
\label{fig:curveiv}
\end{figure}

Figure \ref{fig:curveiv} shows the I-V characteristic curves for the
SiPM of 12 $\mu$m pixel size. The current flowing in the photosensor was measured as a function of the voltage for several doses: the device shows an increase of the dark current by a factor $\sim$300 at the maximum dose (1.7$\times$10$^{11}$ n/cm$^2$) at 32 V, corresponding to an overvoltage of 3.8~V.  The relative increase of the leakage current with the dose for the same SiPM does not depend significantly  on the overvoltage. Moreover, the breakdown edge (at 28.2~V before irradiation) is smoother after the exposure but the change on the breakdown voltage is less then 100~mV.  Waveforms were recorded at each irradiation step and we observed that the sensitivity to single photoelectron is lost at a fluence larger than $\sim$ 1$\times$10$^{10}$ n/cm$^2$.

Several boards hosting the ENUBET SiPMs were irradiated at LNL and have been tested in July and October at CERN-PS (T9 beamline). The boards were installed on the basic calorimeter module of ENUBET (UCM). For these tests we assembled UCMs with different scintillator thicknesses employing a polystyrene-based scintillator produced by Uniplast. The results confirm the possibility to use the SiPMs for the light readout of ENUBET in a broad range of bias voltages and scintillator thicknesses, not only for electron/pion separation but also for the identification of minimum ionizing particles. The mip peak is well separated from the dark noise pedestal if the scintillator thickness is larger than 1 cm. For a scintillator tile thickness of 1.3 cm, the calorimeter provides a clear mip signature up to an integrated neutron fluence of 2$\times$10$^{11}$ n/cm$^2$ (1 MeV-eq). Since the breakdown voltage is not significantly changed by neutron irradiation, the gain and photon detection efficiency (PDE) drop of the SiPM is recovered by increasing the bias voltage in order to equalize the sensor response during the  run. 

\section{Polysiloxane scintillators}
\begin{figure}[htb]
\centering
\includegraphics[height=2.6in]{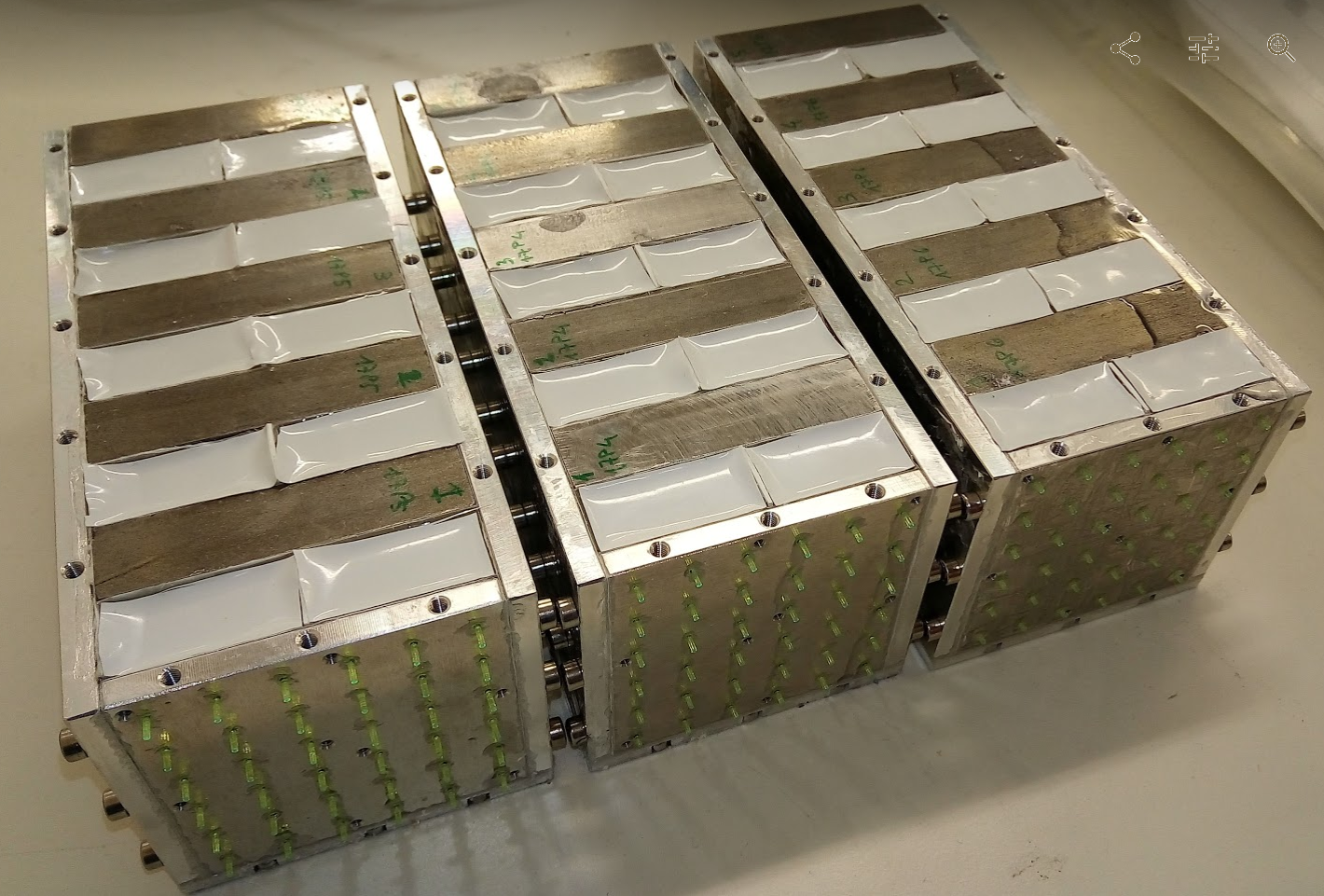}
\caption{Three polysiloxane based modules tested at CERN. The picture is taken before the installation of the SiPM boards.}
\label{fig:siliconico}
\end{figure}

Besides the standard approach that employs plastic scintillator as
active material, alternative solutions have also been considered: in 2017, we built for the first time a shashlik calorimeter that employs a polysiloxane scintillator (1.5 cm Fe as absorber, 1.5 cm scintillator tiles, see Figure \ref{fig:siliconico}). This is a recently developed siliconic-based scintillator~\cite{siliconic1,siliconic2} that offers several advantages over plastic scintillators: better radiation tolerance, reduced aging, no irreversible deterioration caused by
mechanical deformations, exposure to solvent vapours and high temperatures and, for shashlik calorimeters, no need to drill and insert the optical fibers.  Silicone rubbers preserve their transparency even after a 10 kGy dose exposure and their physical properties are constant over a wide temperature range. Polysiloxane can be poured at 60$^\circ$C in the shashlik module after the insertion of the optical fibers in the absorber. Pouring greatly simplifies the process of fiber-insertion and provides an optimal fiber-scintillator optical interface. On the other hand, the light yield is about 30$\%$ of the EJ-200 yield. The performance of the Polysiloxane calorimeter prototype has been assessed during a test beam campaign in October 2017 at the CERN East Area (T9 beamline) \cite{in_preparation}. We measured an electromagnetic energy resolution of 17$\%$/$\sqrt{E(GeV)}$ and a good linearity ($<$ 3$\%$ in the 1-5 GeV range). The light yield per unit thickness is about one-third of EJ-200. As a consequence the quality of the fiber-scintillator coupling after pouring is comparable to the one that can be obtained from injection molding of conventional scintillators. 

\end{document}

%% file: econfmacros.tex
\def\beq{\begin{equation}}
\def\eeq#1{\label{#1}\end{equation}}
\def\eeqn{\end{equation}}

\def\beqa{\begin{eqnarray}}
\def\eeqa#1{\label{#1}\end{eqnarray}}
\def\eeqan{\end{eqnarray}}

\let\bar=\overbar

\def\Dslash{\not{\hbox{\kern-4pt $D$}}}
\def\dslash{\not{\hbox{\kern-2pt $\del$}}}

\def\msb{{\bar{\ssstyle M \kern -1pt S}}}

